\documentclass[aps,pra,12pt,reprint,a4paper,longbibliography,nofootinbib]{revtex4-1}

\usepackage{amsmath,amsthm,amsfonts,amssymb,bm}
\usepackage{graphicx,array,MnSymbol}
\usepackage[usenames,dvipsnames]{xcolor}
\definecolor{OceanBlue}{rgb}{0,0.35,0.7} 
\usepackage{hyperref}
\hypersetup{colorlinks=true,allcolors=OceanBlue}

\theoremstyle{definition}
\theoremstyle{remark}
\newcommand{\pars}[1]{\left(#1\right)} % Parentheses
\newcommand{\bracs}[1]{\left[#1\right]} % Brackets
\newcommand{\lp}{\left(}  % Left paranthesis
\newcommand{\rp}{\right)} % Right paranthesis
\newcommand{\mr}{\mathrm} % Text in math mode
\newcommand {\cl}{\mathcal} % Calligraphic math
\DeclareMathOperator{\Tr}{Tr\,}            % Trace
\DeclareMathOperator{\sech}{sech}   % Hyperbolic secant
\DeclareMathOperator{\csch}{csch}   % Hyperbolic cosecant

\newcommand*\diff{\mathop{}\!\mathrm{d}}
\DeclareMathOperator{\trans}{\textsf{T}} % Transpose

\newcommand {\ket}[1] {\left|{#1}\right\rangle}
\newcommand {\cket}[1] {\left|{#1}\right)}  % Curvy ket

\newcommand {\bra}[1] {\langle{#1}|}
\newcommand {\cbra}[1] {\left({#1}\right|}  % Curvy bra

\newcommand{\cbraket}[2]{\left({#1}|{#2}\right\rangle} % Curvy bra, regular ket

\newcommand{\abs}[1]{\left | #1 \right |}   % Absolute value
\newcommand {\norm} [1] {\left \| #1 \right \|} % Norm
\newcommand{\udl}[1]{\underline{#1}}  % Underline
\begin{document}
\title{Nonclassical distance in multimode bosonic systems}
\author{Ranjith Nair}
\email{elernair@nus.edu.sg}
\affiliation{Department of Electrical and Computer Engineering, \\ National University of Singapore, 4 Engineering Drive 3, 117583 Singapore\\
%Department of Physics, 2 Science Drive 3,
%National University of Singapore, Singapore 117551
}
\date{\today}
\begin{abstract} 
We revisit the notion of nonclassical distance of states of bosonic quantum systems introduced in [\href{http://journals.aps.org/pra/abstract/10.1103/PhysRevA.35.725}{M. Hillery, Phys.~Rev.~A 35, 725 (1987)}] in a general multimode setting. After reviewing its definition, we establish some of its general properties. We obtain new upper and lower bounds on the nonclassical distance in terms of the supremum of the Husimi function of the state. Considering several examples, we elucidate the cases for which our lower bound is tight, which include the multimode number states and a class of multimode N00N states. The latter provide  examples of  states of definite photon number $n \geq 2$ whose nonclassical distance can be made arbitrarily close to the upper limit of $1$ by increasing the number of modes. We show that the nonclassical distance of the even and odd Schr{\"o}dinger cat states is bounded away from unity regardless of how macroscopic the superpositions are, and that the nonclassical distance is not necessarily monotonically increasing with respect to macroscopicity.
\end{abstract}
\maketitle

\section{Introduction}

Consider $M$ independent bosonic
systems such as electromagnetic modes or nanomechanical oscillators and let $\cl{H}$ be the Hilbert space describing the entire system. Let  $S = \cl{S(H)}$ be the system's state space, i.e., the set of all density operators on $\cl{H}$. A distinguished subset of the state space $S$ is the set  $S_{\mr{cl}}$ of \emph{classical states}. For  $\cket{\udl{\alpha}} = \cket{\alpha_1}\cket{\alpha_2} \cdots \cket{\alpha_M},\, \udl{\alpha} \in \mathbb{C}^M$ an $M$-mode coherent state\footnote{For clarity, in this paper we use curved bras and kets \unexpanded{$\cbra{\udl{\alpha}}, \cket{\udl{\beta}}$}, etc. to  indicate coherent states, and the usual bra and ket symbols \unexpanded{$\bra{\phi}, \ket{\psi},$} etc. for general states. For \unexpanded{$n = 0, 1, 2, \ldots$, $\ket{n}$} denotes the number state of \unexpanded{$n$} photons.}, $S_{\rm cl}$ consists of the states $\sigma$ that have a non-negative diagonal-coherent-state or $P$ representation \cite{Sud63,*Gla63,MW95}:-
\begin{align} \label{eq:clstate}
\sigma = \int_{\mathbb{C}^M} \diff^{2M} \udl{\alpha} \;P\pars{\udl{\alpha}}\, \cket{\udl{\alpha}}\cbra{\udl{\alpha}},
\end{align}
where $P\pars{\udl{\alpha}} \geq 0$ is a  probability distribution, possibly with delta-function singularities. We will use the language of quantum optics in this paper, but many of the ideas have counterparts in other continuous-variable systems with effective harmonic-oscillator Hamiltonians. Classical states are important in quantum optics as models for the radiation from commonly-occurring natural sources. Their photodetection statistics can be described in semiclassical terms \cite{MW95}, and they are  readily generated in the laboratory from  laser sources. 

Many characteristic quantum-optical phenomena such as sub-Poissonian photon statistics, antibunching,  quadrature squeezing, and entanglement are displayed only by \emph{nonclassical} states, i.e., states  $\rho \in S_{\rm{ncl}} = S\, \backslash\, S_{\rm{cl}}$. Several nonclassical states have been introduced theoretically and generated experimentally for various applications, well-known examples being the number states of one or more photons, the single-mode and two-mode squeezed coherent states, the Schr\"{o}dinger cat states, and the N00N states -- see, e.g., refs.~\cite{Dod02,Agr12Quantum} and references therein for these and many other examples.  Since the absence of any one of the above characteristic quantum features is not a guarantee that a given state is classical, much attention has been focused on  operational necessary criteria  for nonclassicality -- see, e.g., refs.~\cite{RV02,MBW+10,SCE+11,RSA+15}.

Applications like quantum cryptography, quantum computation, and quantum metrology  rely on nonclassical states to gain a quantum advantage not achievable using classical states. In view of the fact that classical states are much easier to generate than tailored nonclassical ones, and that the latter are very sensitive to decoherence, a quantitative measure of the nonclassicality of a given quantum state is very useful.
Perhaps the earliest such measure proposed is the \emph{nonclassical distance}  introduced by Hillery in  ref.~\cite{Hil87}. It is defined as the minimum trace-norm distance between the given state and states in the set $S_{\rm cl}$, and provides an upper bound on the Kolmogorov (or $l_1$-) distance between the probability distributions obtained on measuring the given state and an arbitrary classical state using \emph{any} quantum measurement. As such, it provides a generic bound on the advantage that the given state can provide over classical ones in any task of interest. Unfortunately, calculating the nonclassical distance is a difficult problem in general, and to the best of our knowledge, no exact results have appeared in the literature. However, several upper and lower bounds on it have been given \cite{Hil87,Hil89}.

Many other measures of nonclassicality have since been defined in the literature. In ref.~\cite{Lee91}, Lee introduced the notion of \emph{nonclassical depth} of a single-mode state, defined as the minimum number of additive Gaussian noise photons required to render the state classical. Techniques for calculating it were also given \cite{Lee92}. The same concept was independently  defined by  L\"utkenhaus and Barnett \cite{LB95}.   While the nonclassical depth is an informative measure of nonclassicality \cite{Lee91} for  Gaussian states \cite{WPG+12}, it was shown in \cite{LB95} that it has the maximal value of 1 noise photon for \emph{any} pure non-Gaussian nonclassical state, rendering it of limited value as a nonclassicality measure for such states. Very recently, Sabapathy has generalized the nonclassical depth to multimode states and also to quantum channels \cite{Sab16}.

Partly motivated by the difficulties of calculating the trace-norm distance (and hence the nonclassical distance), several other distance-based measures of nonclassicality have been proposed. Dodonov \emph{et al.} \cite{DMM+00} introduced the Hilbert-Schmidt distance between a given state and the set of pure coherent states as a measure of nonclassicality. Marian \emph{et al.} \cite{MMS02} studied the minimum Bures distance \cite{Bur69} (closely related to the quantum fidelity \cite{NC00}) between single-mode Gaussian states and the set of classical Gaussian states. Malbouisson and Baseia \cite{MB03} studied the Bures distance and the Hilbert-Schmidt distance of more general states relative to the set of pure coherent states. Marian \emph{et al.} \cite{MMS04} studied the nonclassicality of single-mode Gaussian states using the minimum relative entropy to the set of classical Gaussian states as a `distance' measure.  A measure defined using the Wehrl entropy  has recently been  explored by Bose \cite{Bos17}.  Unfortunately,  in these works, the chosen distance measure has been minimized over only a subset of $S_{\rm cl}$ chosen in a more or less \emph{ad hoc} manner.

In ref.~\cite{ACR05}, Asb\'oth \emph{et al.} exploited the close connection between input nonclassicality and generation of entanglement at the output of a passive linear optics network  to define and estimate new nonclassicality measures.  Vogel and co-workers have used the minimum number of terms in an expansion of the given state as a superposition of coherent states to define an algebraic nonclassicality measure \cite{GSV12,*VS14,*RSV17}. Nonclassicality and entanglement are notoriously fragile under the action of decohering channels, the attenuator, additive Gaussian noise, and amplifier channels \cite{WPG+12,Hol12} being particularly ubiquitous in applications. The degradation of nonclassicality and entanglement under loss and additive noise  has also been extensively studied -- see, e.g., refs.~\cite{ACR05,SVL06,YN09,SIS11}.

In this paper, we revisit the nonclassical distance of ref.~\cite{Hil87} in a general multimode setting.  In Section~\ref{sec:ncldist}, we motivate and review its definition. In Section \ref{sec:properties}, we establish a number of its general properties. In Section \ref{sec:bounds}, we establish improved upper and lower bounds on it in terms of the Husimi function of the given state. In Section \ref{sec:examples}, we consider several examples illustrating our results. In particular, we show that our lower bound is saturated for multimode number states and a class of multimode N00N states, and very nearly saturated for even and odd Schr\"odinger cat states in the ``macroscopic'' regime. We conclude by discussing several possible directions for future work in Section~\ref{sec:discussion}.

\section{Nonclassical distance} \label{sec:ncldist}

Given an arbitrary state $\rho \in S$, its nonclassical distance  $\delta(\rho)$ is defined as
\begin{align} \label{eq:ncldistdef}
\delta(\rho) := \inf_{\sigma \in S_{\mr{cl}}} D(\rho, \sigma) = \frac{1}{2} \inf_{\sigma \in S_{\mr{cl}}} \norm{\rho - \sigma}_1,
\end{align}
where $D(\rho, \sigma)$ is the trace distance between $\rho$ and $\sigma$, with the latter  of the form of Eq.~\eqref{eq:clstate}.  $\norm{X}_1 := \Tr \sqrt{X^\dag X}$ is the trace norm of the trace-class operator $X$. Note that our definition differs from that in \cite{Hil87} by a factor of $1/2$. Following a convention often used in quantum information \cite{NC00}, we have used the trace distance to measure the separation between $\rho$ and $S_{\rm cl}$ -- the nonclassical distance then satisfies $0 \leq \delta(\rho) \leq 1$. A classical state evidently has zero nonclassical distance. We show later that $\delta(\rho) <1$ for all $\rho \in S$ (see Sec.~\ref{sec:ubounds}). In Sec.~\ref{sec:positivity}, we show that $\delta(\rho) >0$ for all $\rho \in S_{\rm ncl}$ so that $\delta(\rho) > 0$ is a necessary and sufficient condition for nonclassicality. 

To appreciate the utility of the definition \eqref{eq:ncldistdef}, consider an arbitrary positive operator-valued measure (POVM) \cite{NC00}  $\{\Pi(x)\}_{x \in \cl{X}}$ describing a quantum measurement yielding an outcome $x$ in an arbitrary measurable space $\cl{X}$ and  let $\sigma \in S_{\rm cl}$. The Kolmogorov distance between the  classical probability densities $P_\rho(x) = \Tr \rho \Pi(x) $ and $P_\sigma(x) = \Tr \sigma \Pi(x)$ resulting from measuring the given POVM on $\rho$ and $\sigma$ is 
\begin{align}
K\pars{P_\rho, P_\sigma} &:= \frac{1}{2}\int_\cl{X} \diff x \abs{P_\rho(x) - P_\sigma(x)} \\
&=  \frac{1}{2}\int_\cl{X} \diff x \abs{\Tr (\rho - \sigma) \Pi(x) } \\
&\leq \frac{1}{2}\int_\cl{X} \diff x \Tr \abs{\rho - \sigma} \Pi(x)  \\
&=  \frac{1}{2} \norm{\rho - \sigma}_1,
\end{align}
where  the inequality follows from the fact that $X \leq \abs{X} = \sqrt{X^\dag X}$ for Hermitian and trace-class $X$ and that $\Pi(x) \geq 0$. As is well known, the inequality is saturated by the Helstrom-Holevo measurement \cite{Hel76,Hol73}. It follows that
\begin{align}
\inf_{\sigma \in S_{\rm cl}} \max_{\{\Pi_x\}} K\pars{P_\rho, P_\sigma} = \inf_{\sigma \in S_{\rm cl}} D(\rho, \sigma) = \delta(\rho).
\end{align}
The nonclassical distance thus provides a measurement-independent (and hence, application-independent) quantification of the minimum distinguishability between the probability distributions generated by the given state and any classical state. In other words, it generically quantifies the possible advantage that can be gained from the given nonclassical state (which is an  ``expensive'' resource in being typically hard to generate and preserve) over classical states (which are ``cheap'' resources as they are easy to generate). Unfortunately, the nonclassical distance is difficult to calculate in general, owing both to the difficulty of calculating the trace distance in Eq.~\eqref{eq:ncldistdef} (which requires diagonalizing $\rho - \sigma$)  and to having to minimize it over all of $S_{\rm cl}$. Both these difficulties are compounded by the infinite dimensionality of $\cl{H}$.

%\newpage
\section{General Properties of $\delta(\rho)$} \label{sec:properties}
We now establish a number of useful properties of $\delta(\rho)$, many of which are used in the examples to follow. Most of the properties are intuitive, but (unless indicated) have not been formally stated in the literature to the best of our knowledge.  Some of the results -- such as the positivity of $\delta(\rho)$ for all $\rho \in S_{\rm ncl}$ -- are not obvious.

\subsection{Positivity for $\rho \in S_{\rm ncl}$} \label{sec:positivity}
 $S_{\rm cl}$ has an interesting topological structure within $S$. Using an ingenious argument, it was shown in \cite{Hil87} that  there are nonclassical states arbitrarily close in trace distance to \emph{any} single-mode classical state. This argument is readily adapted to the multimode case, showing that $S_{\rm ncl}$ is dense in $S$  with respect to the trace distance.
On the other hand, it was also shown in \cite{Hil87} that the non-vacuum number states (which are nonclassical\footnote{From the definition \eqref{eq:clstate}, we see that a non-vacuum classical state must have a positive probability of counting any number of photons in some mode. Thus, the non-vacuum multimode number states are nonclassical.}), are interior points of $S_{\rm ncl}$ with positive nonclassical distance. In order to generalize this statement to all $\rho \in S_{\rm ncl}$, we note that the convex set $S_{\rm cl}$ is  closed with respect to the weak topology on $S$ \cite{BL86} and is therefore also closed with respect to trace distance (\cite{Hol12}, Lemma 11.1). Equivalently, $S_{\rm ncl}$ is open in $S$ so that the nonclassical distance is strictly positive for {all} $\rho \in S_{\rm ncl}$. Positivity of $\delta(\rho)$ is thus a necessary and sufficient condition for $\rho$ to be nonclassical.

This juxtaposition of  $S_{\rm cl}$ and $S_{\rm ncl}$  is analagous to that of the set of separable states and the set of non-separable states of a bipartite system of which at least one subsystem is infinite-dimensional. For such systems, it was shown in \cite{CH99} that the set of nonseparable states is open and dense in the set of all states with respect to trace distance.

\subsection{Non-increase under classicality-preserving channels} \label{sec:clpchannels}
Let $T: S \rightarrow S'$ be a quantum channel (a completely positive trace-preserving map) \cite{NC00,Hol12} between the (not necessarily identical) state spaces $S$ and $S'$ that takes classical states to classical states, i.e., $T \,S_{\rm cl} \subset S'_{\rm cl}$. Such channels are exactly the ones defined as ``classical channels'' in \cite{R-KKV+13} and include the more restrictive nonclassicality-breaking channels \cite{Sab16}. We have
\begin{align}
\delta(T\rho) &= \inf_{\sigma^\prime \in S'_{\rm cl}} D(T\rho, \sigma') \\
& \leq \inf_{\sigma \in S_{\rm cl}} D(T\rho, T\sigma) \\
& \leq \inf_{\sigma \in S_{\rm cl}} D(\rho, \sigma) \\
&= \delta(\rho),
\end{align}
where the first inequality follows from the assumption that $T$ is classicality-preserving and the second  follows from the non-increase of trace distance under the action of quantum channels.

\subsection{ Invariance to adjoining a classical state} \label{sec:adjclstate}
 Suppose we have two subsystems $A$ and $B$ each consisting of one or more modes. Let $\rho \in S^A$ be a state of $A$ and $\sigma_0 \in S_{\rm cl}^B$ be a classical state of $B$. We have
\begin{align} \label{eq:invadj}
\delta(\rho \otimes \sigma_0) = \delta(\rho).
\end{align}
To see this, first note that
\begin{align}
\delta(\rho \otimes \sigma_0)  \geq \delta(\rho)
\end{align}
from Sec.~\ref{sec:clpchannels} since taking the partial trace over $B$ is a classicality-preserving quantum channel. On the other hand, if $\sigma^A \in S_{\rm cl}^A$, $\sigma^A \otimes \sigma_0 \in S_{\rm cl}^{AB}$, and we have
\begin{align}
\norm{\rho \otimes \sigma_0 - \sigma^A \otimes \sigma_0 }_1 = \norm{\rho - \sigma^A  }_1,
\end{align}
so that
\begin{align}
\delta(\rho \otimes \sigma_0) &\leq \frac{1}{2}
\inf_{\sigma^A \in S_{\rm cl}^{A}} \norm{\rho \otimes \sigma_0 - \sigma^A \otimes \sigma_0 }_1 \\
&= \delta(\rho).
\end{align}
Thus, adjoining a classical state does not change the nonclassical distance, as may be expected intuitively. 

\subsection{ Invariance under affine optics  transformations}  \label{sec:aoptics}
Let $\udl{a} = (a_1,\ldots, a_M) ^{\trans}$  be the  vector of annihilation operators corresponding to the $M$ modes of the system.  Consider the Heisenberg-picture unitary transformation
\begin{align} \label{eq:HPunitary}
\udl{b} = U\udl{a} + \udl{\gamma},
\end{align}
where $\udl{b} = (b_1,\ldots, b_M)^{\trans}$ is the  output vector of annihilation operators, $U$ is an $M \times M$ unitary matrix, and $\udl{\gamma} = (\gamma_1, \ldots, \gamma_M)^{\trans}$ is an arbitrary vector in $\mathbb{C}^M$. Such a transformation corresponds to the most general unitary transformation that can be performed on the input modes using passive linear optics, augmented by displacements in each mode by amounts given by $\udl{\gamma}$  (hence the terminology ``affine optics transformation''). The corresponding quantum channel $T_{U,\udl{\gamma}}$ maps (in the Schr\"odinger picture) product coherent states into product coherent states according to
\begin{align} \label{eq:SPunitary}
\cket{\udl{\alpha}} \mapsto \cket{ U\udl{\alpha} + \udl{\gamma}}.
\end{align}
We thus have $T_{U,\udl{\gamma}} S_{\rm cl} = S_{\rm cl}$ so that for any $\rho \in S$,
\begin{align}
\delta(T_{U,\udl{\gamma}}\rho) &= \inf_{\sigma \in S_{\rm cl}} D(T_{U,\udl{\gamma}}\;\rho, \sigma) \\
& =\inf_{\sigma \in S_{\rm cl}} D(T_{U,\udl{\gamma}}\;\rho, T_{U,\udl{\gamma}}\;\sigma) \\
& =\inf_{\sigma \in S_{\rm cl}} D(\rho, \sigma) \\
&= \delta(\rho),
\end{align}
where we have used the unitary invariance of the trace distance.

The above result may be viewed as a quantitative generalization to $M$-mode affine optics transformations of the well-known fact that 2-port beamsplitters cannot generate nonclassical states from classical ones \cite{AM99,Wan02}.  A similar invariance result for the Bures distance has been stated by Marian \emph{et al.} \cite{MMS04}, and the special case of the above result for pure dispacements was shown in ref.~\cite{Hil87}.  The preceding two properties imply that nonlinear processes of second order or higher are required to increase the nonclassical distance of a given state  augmented by auxiliary modes in classical states.

\subsection{ Convexity} \label{sec:convexity}
The nonclassical distance is convex in $\rho$, as shown in ref.~\cite{Hil89}. Indeed, let $\rho_1,\rho_2 \in S$,  $\sigma_1,\sigma_2 \in S_{\rm cl}$, and $\rho = (1-\epsilon) \rho_1 + \epsilon \rho_2$  for any $\epsilon$ with $0 \leq \epsilon \leq 1$. The state $ \sigma = (1-\epsilon) \sigma_1 + \epsilon \sigma_2 \in S_{\rm cl}$ and we  have
\begin{align}
\norm{\rho - \sigma}_1 \leq (1-\epsilon) \norm{\rho_1 - \sigma_1}_1 + \epsilon \norm{\rho_2 - \sigma_2}_1
\end{align}
by convexity of the trace norm, so that
\begin{align}
&\inf_{\sigma' \in S_{\rm cl}} \norm{\rho - \sigma'}_1 \\
&\leq (1-\epsilon) \inf_{\sigma_1 \in S_{\rm cl}}\norm{\rho_1 - \sigma_1}_1 + \epsilon \inf_{\sigma_2 \in S_{\rm cl}}\norm{\rho_2 - \sigma_2}_1,
\end{align}
and so
\begin{align}
\delta(\rho) \leq (1-\epsilon)\,\delta(\rho_1) + \epsilon\,\delta(\rho_2).
\end{align}

\section{Upper and lower bounds on the nonclassical distance} \label{sec:bounds}

The Husimi $Q$ function is a quasiprobability distribution that plays a major role in theoretical quantum optics \cite{Agr12Quantum,Hus40,*Kan65,*MS65,*CG69b} and is also experimentally accessible via heterodyne detection \cite{Leo97measuring}. For given $\rho \in S$ and $\udl{\alpha} \in \mathbb{C}^M$, it is defined as
\begin{align}
Q_{\rho}(\udl{\alpha}) =  \cbra{\udl{\alpha}}\rho \cket{\udl{\alpha}}/\pi^M
\end{align}
and is a normalized true probability density. Some of the bounds on the nonclassical distance
derived in refs.~\cite{Hil87,Hil89} involve the supremum of the Husimi function of the state.  For a general state $\rho \in S$, let us define
\begin{align}
\widetilde{Q}_\rho\pars{\udl{\alpha}} &:=  \cbra{\udl{\alpha}} \rho \cket{\udl{\alpha}} = \pi^M Q_{\rho}(\udl{\alpha}); \\
m(\rho) &:= \sup_{\udl{\alpha} \in \mathbb{C}^M} \widetilde{Q}_{\rho}(\udl{\alpha}). \label{eq:mdef}
\end{align}
In this Section, we make use of relations between the quantum fidelity and trace distance \cite{FvdG99,NC00} in order to obtain stronger upper and lower bounds on $\delta(\rho)$ in terms of $m\pars{\rho}$ than those in ref.~\cite{Hil87}. These relations also allow for more direct derivations and make the cases for which the bounds are saturated more transparent.

\subsection{Lower bounds} \label{sec:lbounds}
The  fidelity  between any two states $\rho$ and $\sigma$ is given by $F(\rho, \sigma) = \Tr \sqrt{\sqrt{\rho} \sigma \sqrt{\rho}}$ \cite{NC00}. The trace distance and fidelity obey the inequality  \cite{FvdG99}:-   
\begin{align} \label{eq:DlbinF}
1-F(\rho, \sigma) \leq D(\rho,\sigma).
\end{align}
If $\rho \in S$ is given and  $\sigma \in S_{\mr{cl}}$ is a classical state of the form \eqref{eq:clstate}, this gives the lower bound
\begin{align} \label{eq:mixedstatelb}
1 - \sup_{\sigma \in S_{\rm{cl}}} F(\rho, \sigma) \leq \delta(\rho)
\end{align}
on the nonclassical distance of $\rho$. Since fidelity is a concave function of its arguments \cite{NC00}, this expression does not readily simplify on substituting Eq.~\eqref{eq:clstate} for $\sigma$.

For \emph{pure} states $\rho = \ket{\psi} \bra{\psi}$, a stronger inequality than \eqref{eq:DlbinF} holds \cite{NC00,AM14}\footnote{We use expressions like \unexpanded{$\delta(\ket{\psi}), F(\ket{\psi}, \sigma), D(\ket{\psi}, \sigma),$} etc. with the obvious meanings when pure states are involved.}:
\begin{align} \label{eq:fidlbonD}
1-F^2(\ket{\psi}, \sigma) = 1 -\bra{\psi}\sigma\ket{\psi} \leq D(\ket{\psi},\sigma).
\end{align}
 For $\sigma$ of the form of Eq.~\eqref{eq:clstate}, this gives the lower bound
\begin{align} \label{eq:purestlb1}
\delta(\ket{\psi}) &\geq 1-  \sup_{P\pars{\udl{\alpha}}}\int_{\mathbb{C}^M} \diff^{2M} \udl{\alpha} \;P\pars{\udl{\alpha}}\,\abs{\cbra{\udl{\alpha}}\psi \rangle}^2     \\
&=  1-  m\pars{\ket{\psi}}, \label{eq:purestlb} 
\end{align}
since $P(\udl{\alpha}) \geq 0$ and integrates to one. This strengthens (by a factor of 2)  the bound (4.2) of \cite{Hil87} specialized to pure states. 

Since $m\pars{\otimes_{m=1}^M\ket{\psi}_m} = \Pi_{m=1}^M m\pars{\ket{\psi_m}}$, we have that
\begin{align}
\delta\pars{\otimes_{m=1}^M\ket{\psi_m}} \geq 1 - \Pi_{m=1}^M m\pars{\ket{\psi_m}}.
\end{align}
In particular, the nonclassical distance of a product of identical pure nonclassical states approaches 1 at least exponentially fast in the number of copies.

\subsection{Upper bounds} \label{sec:ubounds}
We can also obtain upper bounds on $\delta(\rho)$ via the fidelity. Using the inequality \cite{FvdG99}
\begin{align} \label{eq:DubviaF}
 D^2(\rho,\sigma) \leq 1 -  F^2(\rho,\sigma),
\end{align}
we have for any classical state $\sigma$ of the form of Eq.~\eqref{eq:clstate},
\begin{align}
D^2(\rho,\sigma) 
&\leq 1 - \bracs{F\pars{\rho, \int_{\mathbb{C}^M} \diff^{2M} \udl{\alpha} \;P\pars{\udl{\alpha}}\, \cket{\udl{\alpha}}\cbra{\udl{\alpha}}}}^2 \\
&\leq 1- \bracs{\int_{\mathbb{C}^M} \diff^{2M} \udl{\alpha} \;P\pars{\udl{\alpha}}\,F\pars{\rho,  \cket{\udl{\alpha}}\cbra{\udl{\alpha}}}}^2, 
\end{align}
using the concavity of fidelity. It follows that
\begin{align}
\delta(\rho) &= \inf_{P(\udl{\alpha})} D(\rho, \sigma) \\
& \leq \left\{1 -\sup_{P(\udl{\alpha})} \bracs{\int_{\mathbb{C}^M} \diff^{2M} \udl{\alpha} \;P\pars{\udl{\alpha}}\,F\pars{\rho,  \cket{\udl{\alpha}}\cbra{\udl{\alpha}}} }^2\right\}^{1/2} \\
&= \bracs{1 - m(\rho)}^{1/2}. \label{eq:genstateub}
\end{align}
This  result is a generalization of the bound (4.14) of \cite{Hil87} to mixed states. 

Since $m(\rho) > 0$ (Otherwise, $Q_\rho(\udl{\alpha}) \equiv 0$, which is impossible since $\int_{\mathbb{C}^M} \diff^{2M} \udl{\alpha}\; Q_\rho(\udl{\alpha}) =1$), the upper bound shows that $\delta(\rho) <1$ for all states.

 If $\rho = \ket{\psi}\bra{\psi}$ is pure \emph{and} the closest classical state (assuming one exists) to $\ket{\psi}$ is a pure coherent state, the above upper bound is an equality because all the inequalities from \eqref{eq:DubviaF} to \eqref{eq:genstateub} are saturated in this case. A coherent state $\cket{\udl{\alpha}_\star}$ that satisfies $m\pars{\ket{\psi}} = \abs{\cbraket{\udl{\alpha}_\star}{\psi}}^2$ indeed achieves the greatest possible fidelity among $\sigma \in S_{\rm cl}$ (and hence the smallest possible Bures distance) with a given $\ket{\psi} \in S_{\rm ncl}$, as the case of saturation of \eqref{eq:purestlb} shows\footnote{Thus, for pure states \unexpanded{$\ket{\psi}$}, the lower bound \eqref{eq:purestlb} in fact determines the minimum Bures distance with respect to \unexpanded{$S_{\rm cl}$}, making the restriction to the pure states of \unexpanded{$S_{\rm cl}$} imposed in \cite{MB03} unnecessary.}. However, it is \emph{not} the case in general that a pure coherent state is the closest classical state \emph{in trace distance} to a given pure nonclassical state, as we will see in Section \ref{sec:examples}.

Overall, for pure states $\ket{\psi}$, we thus have the two-sided $Q$-function-based bounds
\begin{align} \label{eq:2smb}
1 - m\pars{\ket{\psi}} \leq \delta\pars{\ket{\psi}} \leq \bracs{1 - m\pars{\ket{\psi}}}^{1/2}.
\end{align}

Using the triangle inequality for the trace distance, we can relate the nonclassical distances of two states $\rho$ and $\rho'$ as follows. For any $\sigma \in S_{\rm cl}$, we have
\begin{align} 
D(\rho',\sigma) \leq D(\rho,\rho') + D(\rho,\sigma),
\end{align}
so that
\begin{align} \label{eq:tribound}
\abs{\delta(\rho) - \delta(\rho')} \leq D(\rho,\rho'),
\end{align}
giving an upper (lower) bound on the larger (smaller) of $\delta(\rho)$ and $\delta(\rho')$.

\section{Examples} \label{sec:examples}
We now illustrate our general results of Secs.~\ref{sec:properties}-\ref{sec:bounds} with a few examples, and also obtain the exact value of the nonclassical distance for some states.

\subsection{Multimode number states}
First consider the single-mode number states $\ket{n},\,n=0,1,2,\ldots$  Following ref.~\cite{Hil87}, let us define the numbers
\begin{align}
\gamma_n &:= \sup_{\alpha \in \mathbb{C}} \abs{\cbraket{\alpha}{n}}^2 \\
&= \sup_{x \geq 0} e^{-x}\frac{x^n}{n!} \\
&=  \left\{
	\begin{array}{ll}
		1  & \mbox{if } n=0,  \\
		 e^{-n}\frac{n^n}{n!}& \mbox{otherwise.}
	\end{array}
\right.
\end{align}
$\gamma_n$ is thus the probability that a Poisson random variable of mean $n$ takes the value $n$.  It can be verified that $\{\gamma_n\}$ is a decreasing sequence and that it satisfies
\begin{align}
e^{-n} \leq \gamma_n \leq \frac{1}{\sqrt{2\pi n}},
\end{align}
where the upper bound follows from Stirling's approximation for $n!$ and is the asymptote of $\gamma_n$ for large $n$. The lower bound \eqref{eq:purestlb} for the number state $\ket{n}$ is then
\begin{align} \label{eq:nlb}
\delta(\ket{n}) \geq 1 -\gamma_n.
\end{align}
Consider the classical state 
\begin{align}
\overset{\circ}{\sigma}_n &= \frac{1}{2\pi}\int_0^{2\pi} \diff \theta\cket{\sqrt{n}\,e^{i\theta}}\cbra{\sqrt{n}\,e^{i\theta}}  \label{eq:prcs}\\
&= e^{-n}\sum_{m=0}^\infty \frac{n^m}{m!} \ket{m}\bra{m}, 
\end{align}
which (as the overcircle on $\sigma$ suggests) is the uniformly phase-randomized coherent state of mean photon number $n$. The states $\overset{\circ}{\sigma}_n$ and $\ket{n}\bra{n}$ commute and the trace distance between them is thus
\begin{align}
D\pars{\overset{\circ}{\sigma}_n, \ket{n}} = 1 -\gamma_n,
\end{align}
saturating the lower bound \eqref{eq:nlb}. 

Now let $\ket{\udl{n}} =\ket{n_1}\ket{n_2} \cdots \ket{n_M}$ be a product $M$-mode number state. We have
\begin{align}
m(\ket{\udl{n}}) = \sup_{\udl{\alpha} \in \mathbb{C}^M} \abs{\cbraket{\udl{\alpha}}{\udl{n}}}^2 = \prod_{m=1}^M \gamma_{n_m},
\end{align}
so that the lower bound
\begin{align} \label{eq:numstatelb}
\delta(\ket{\udl{n}}) \geq 1 - \prod_{m=1}^M \gamma_{n_m}
\end{align}
holds. As for the single-mode case, $\ket{\udl{n}}$ is an eigenstate of the classical state
\begin{align}
\overset{\circ}{\sigma}_{\udl{n}} := \overset{\circ}{\sigma}_{n_1} \otimes \cdots \otimes \overset{\circ}{\sigma}_{n_M}
\end{align} 
with eigenvalue $\prod_{m=1}^M \gamma_{n_m}$, so that
\begin{align}
D\pars{\overset{\circ}{\sigma}_{\udl{n}}, \ket{\udl{n}}} = 1 - \prod_{m=1}^M \gamma_{n_m} =
\delta(\ket{\udl{n}}).
\end{align}
Since $\gamma_n$ decreases with increasing $n$, we see that increasing the photon number in any  mode increases the nonclassical distance, as may be expected. More interestingly, consider the case where the total photon number $n = \sum_{m=1}^M n_m$ is fixed but the number of modes may be varied. We then have
\begin{align}
\delta(\ket{\udl{n}}) = 1 - \prod_{m=1}^M \gamma_{n_m} \leq 1 - \prod_{m=1}^M e^{-n_m} = 1 -e^{-n},
\end{align}
with the maximum achieved by a product state with one photon in each mode. Thus, spreading out the available energy over many modes increases the nonclassical distance of a multimode number state.

\subsection{Superposition states of definite total photon number}
Consider now an  $M$-mode single-photon state
\begin{align} \label{eq:spstate}
\ket{\psi} = \sum_{m=1}^M c_m \underbrace{\ket{0}\cdots \ket{0}}_{\text{$(m-1)$ modes}}\hspace{-1mm}\ket{1}\hspace{-1mm} \underbrace{\ket{0} \cdots \ket{0}}_{\text{$(M-m)$ modes}},
\end{align}
with arbitrary coefficients $\{c_m\}_{1}^M;\, \sum_{m=1}^M \abs{c_m}^2 =1$.  We have, for $\norm{\udl{\alpha}}_2 = \pars{\sum_{m=1}^M \abs{\alpha_m}^2}^{1/2}$  the Euclidean norm on $\mathbb{C}^M$,
\begin{align}
\abs{\cbraket{\udl{\alpha}}{\psi}}^2 &= e^{-\norm{\udl{\alpha}}_2^2}\abs{ \sum_{m=1}^M \alpha_m^* c_m }^2 \\
&\leq  e^{-\norm{\udl{\alpha}}_2^2}  \norm{\udl{\alpha}}_2^2 \pars{\sum_{m=1}^M \abs{c_m}^2} \\
&\leq \gamma_1 = e^{-1},
\end{align}
which is achieved for $\alpha_m = c_m$, so that $1- e^{-1} \leq \delta(\ket{\psi})$. In fact, $\ket{\psi}$ can be obtained from the state $\ket{1}\ket{0}\cdots\ket{0}$ with nonclassical distance $(1 -e^{-1})$  by using a linear passive transformation of the form \eqref{eq:HPunitary} with $\udl{\gamma} =0$ and any unitary $U$ whose first column consists of the coefficients $\{c_m\}_{m=1}^M$. Therefore, using the properties of nonclassical distance from Sec.~\ref{sec:properties}, any state of the form of Eq.~\eqref{eq:spstate} has  nonclassical distance $(1 -e^{-1})$ independent of the coefficients and the number of modes.

The situation for superposition states of two or more photons is very different.  While the general case appears to be complicated,  consider states of the form
\begin{align} \label{eq:npstate}
\ket{\psi} &=\sum_{m=1}^M c_m\hspace{-2mm}\underbrace{\ket{0}\cdots \ket{0}}_{\text{$(m-1)$ modes}}\hspace{-2mm}\ket{n} \hspace{-2mm}\underbrace{\ket{0} \cdots \ket{0}}_{\text{$(M-m)$ modes}}\hspace{-2mm}
\equiv  \sum_{m=1}^M c_m\, \ket{\psi_m},
\end{align}
where  $n \geq 2$, and the coefficients $\{c_m\}$ are arbitrary.  For $\udl{\alpha} \in \mathbb{C}^M$, we have
\begin{align}
\abs{\cbraket{\udl{\alpha}}{\psi}}^2 &= \frac{e^{-\norm{\udl{\alpha}}_2^2}}{{n!}} \abs{\sum_{m=1}^M  {\alpha_m^*}^n\,c_m}^2 \\
&\leq  \frac{e^{-\norm{\udl{\alpha}}_2^2}}{{n!}}  \pars{\sum_{m=1}^M  \abs{\alpha_m}^n}^2 \pars{\max_m \abs{c_m}}^2\\
&=  \frac{e^{-\norm{\udl{\alpha}}_2^2}}{{n!}}  \norm{\udl{\alpha}}_n^{2n} \pars{\max_m \abs{c_m}}^2 \\
&\leq \frac{e^{-\norm{\udl{\alpha}}_2^2}}{{n!}} \norm{\udl{\alpha}}_2^{2n} \pars{\max_m \abs{c_m}}^2.
\end{align}
Here, $\norm{\udl{\alpha}}_p = \pars{\sum_{m=1}^M \abs{\alpha_m}^p}^{1/p}$ is the $l_p$-norm on $\mathbb{C}^M$ and we have used the inequality $\norm{\udl{\alpha}}_p \geq \norm{\udl{\alpha}}_q$  for $p <q$  \cite{HJ12Matrix}. The inequalities above are saturated by choosing $\udl{\alpha}$ = $(0,\ldots, 0, \sqrt{n}\,e^{-i{\theta_{m_*}}/n}, 0,\ldots, 0)$,  where $m_* =\arg\max \abs{c_m}$ and $\theta_m = \angle c_m$, so that
\begin{align}
m(\ket{\psi}) = {\gamma_n}\pars{\max_m \abs{c_m}}^2.
\end{align}
The lower bound \eqref{eq:purestlb} is then maximized by choosing $\abs{c_m} = 1/\sqrt{M}$ for all $m$, in which case
\begin{align}
\delta(\ket{\psi}) \geq 1- \frac{\gamma_n}{M}.
\end{align}
Thus, in contrast to the single-photon case, for any $n\geq2$, the nonclassical distance of the state \eqref{eq:npstate} with $c_m = e^{i\theta_m}/\sqrt{M}$ can be made arbitrarily close to 1 by increasing the number of modes. We call such states \emph{multimode N00N} states, the usual N00N states \cite{Dow09} being recovered for $M=2$ and $c_m =1/\sqrt{2}$.

We can  show that the Husimi-function lower bound is achieved for multimode N00N states. For $m=1,\ldots, M$, let
\begin{align} \label{eq:sigmam}
\sigma^{(m)} = \underbrace{\cket{0}\cbra{0}\otimes \cdots \cket{0}\cbra{0}}_{\text{$(m-1)$ modes}}\otimes \hspace{0.5mm}\overset{\circ}{\sigma}_{n}\otimes\underbrace{\cket{0}\cbra{0}\otimes \cdots \cket{0}\cbra{0}}_{\text{$(M-m)$ modes}},
\end{align}
so that for each term  in the superposition \eqref{eq:npstate},
\begin{align}
&\sigma^{(m)}\ket{\psi_{m'}} = \gamma_n\, \delta_{mm'}\ket{\psi_{m'}}.
\end{align}
The classical state
\begin{align}
\sigma = \frac{1}{M} \sum_{m=1}^M \sigma^{(m)}
\end{align}
satisfies 
\begin{align}
\sigma \ket{\psi} = \frac{\gamma_n}{M} \ket{\psi},
\end{align}
so that 
\begin{align}
D(\sigma, \ket{\psi}) = 1 -\frac{\gamma_n}{M} = \delta\pars{\ket{\psi}}.
\end{align}
In particular, for the N00N state $\ket{\chi_n} = \pars{\ket{n}\ket{0} + e^{i\theta}\,\ket{0}\ket{n}}/\sqrt{2}$ with $n$ photons, we  have
\begin{align} \label{eq:noon}
\delta\pars{\ket{\chi_n}} = \left\{
	\begin{array}{ll}
		1 - \gamma_1  & \mbox{if } n=1,  \\
		 1 -\frac{\gamma_n}{2}& \mbox{otherwise,}
	\end{array}
\right.
\end{align}
which for $n \geq 2$ is greater than that of the single-mode number state $\ket{n}$ and may be seen as a consequence of the entanglement in the former.
Observe that  $\ket{\chi_2}$ has  the nonclassical distance distance $1-e^{-2} = \delta(\ket{1}\ket{1})$, as dictated by linear optics invariance.

\subsection{Attaining the $Q$-function lower bound} \label{sec:attainlb}

The above examples prompt the general question: ``Which pure states $\ket{\psi}$ saturate the lower bound \eqref{eq:purestlb}?'' The derivation of \eqref{eq:purestlb} indicates that for a classical state $\sigma$ to saturate it, two conditions must hold:-- (a) We must  have equality in \eqref{eq:fidlbonD} and (b) we must have $F^2\pars{\ket{\psi},\sigma} = m\pars{\ket{\psi}}$. The derivation of \eqref{eq:fidlbonD} in turn shows that condition (a) holds only if $\ket{\psi}$ is an eigenvector of $\sigma$ \cite{AM14}. Condition (b) holds only if $\sigma$ is a mixture of coherent states that all (other than a set of probability zero)  have the maximum possible overlap $m\pars{\ket{\psi}}$ with $\ket{\psi}$. It can be verified that the examples  considered so far satisfy these conditions.

An important class of states for which the lower bound \emph{cannot} be saturated are the multimode Gaussian pure states \cite{WPG+12}. Indeed, these have Gaussian $Q$ functions that are maximized at exactly one phase-space point --  the  mean vector of the given state. Thus, condition (b) above cannot be satisfied for any pure nonclassical Gaussian state, and hence also the lower bound \eqref{eq:purestlb}. It is an open question if the upper bound in \eqref{eq:2smb} is saturated for these states.

\subsection{Even and Odd Schr\"odinger cat states} \label{sec:escstate}
Consider the even and odd superpositions of the single-mode coherent states $\cket{\pm \beta}$:
\begin{align} \label{eq:scstates}
\ket{\psi_\pm} &= \frac{\cket{\beta} \pm \cket{-\beta}}{\sqrt{2\cl{N_\pm}}}, 
\end{align}
i.e., the \emph{even and odd coherent states} introduced in ref.~\cite{DDM74} (see also \cite{Ger93}). Linear optics invariance implies that in order to examine the nonclassical distance of these states, we can set the amplitude $\beta >0$ without loss of generality. The normalization constants are
\begin{align}
\cl{N_\pm} = 1 \pm e^{-2\beta^2}.
\end{align}  Such states have been generated in microwave cavities \cite{Har13}, in the motional degree of freedom of trapped ions \cite{Win13},  in traveling optical beams \cite{OJT+07,LGC+13,HLJ+15} and other systems \cite{AH14}. For large values of $\beta$, they are an example of the ``Schr\"odinger cat'' states \cite{Sch35} involving superpositions of macroscopic states, and are of great interest in studies of the quantum-classical divide \cite{Leg02}. The nonclassicality of the states \eqref{eq:scstates} has been studied from the point of view of Bures distance in \cite{MB03}.

The $Q$ functions of the states \eqref{eq:scstates} are:
\begin{align} \label{eq:Qfuncscstates}
\widetilde{Q}_{\ket{\psi_+}}({\alpha}) &= (\sech \beta^2)\, e^{-\abs{\alpha}^2} \abs{\cosh \beta \alpha}^2, \\
\widetilde{Q}_{\ket{\psi_-}}({\alpha}) &= (\csch \beta^2)\, e^{-\abs{\alpha}^2} \abs{\sinh \beta \alpha}^2.
\end{align}
Evaluating the bounds \eqref{eq:2smb} requires obtaining the maximum of these functions over  $\alpha \in \mathbb{C}$. 

In finding the maximum of the $Q$ function for the even coherent state, two cases arise. If $\beta \leq 1$, the $Q$ function has a single maximum at the origin, while if $\beta >1$, there are two maxima of equal height at $\pm \alpha_*$ where $0  < \alpha_* < \beta$ is the nonzero solution of  $\beta \tanh (\beta \alpha_*) = \alpha_*$ (see inset to Fig.~1).  For any value of $\beta > 0$, it can be shown that the maxima of the $Q$ function of the odd coherent state occur on the real axis at $\pm \alpha_*$ satisfying $\beta\,\coth(\beta \alpha_*) = \alpha_*$ (see inset to Fig.~2). In the limit $\beta \rightarrow 0, \ket{\psi_-} \rightarrow \ket{1}$, the one-photon state, so that  the $Q$ function is maximized at any coherent state of the form $\cket{e^{i\theta}}$ in that limit.

Consider the classical incoherent mixture
\begin{align} \label{eq:scsigma}
\sigma_\beta &= \frac{1}{2}\cket{\beta}\cbra{\beta}+ \frac{1}{2}\cket{-\beta}\cbra{-\beta}
\end{align}
of the states $\cket{\pm \beta}$. For any value of $\beta$, $\ket{\psi_\pm}$ are eigenvectors of $\sigma_\beta$:
\begin{align} \label{eq:scsigmaev}
\sigma_\beta \ket{\psi_\pm}& = \frac{\cl{N}_\pm}{2} \ket{\psi_\pm},
\end{align}
so that we have the upper bounds
\begin{align} \label{eq:scubb}
\delta(\ket{\psi_\pm}) \leq D(\ket{\psi_\pm},\sigma_\beta) = \frac{1  \mp e^{-2\beta^2}}{2}.
\end{align}
Similarly, the trace distance to the classical state
\begin{align} \label{eq:scuba}
\sigma_{\alpha_*} &= \frac{1}{2}\cket{\alpha_*}\cbra{\alpha_*}+ \frac{1}{2}\cket{-\alpha_*}\cbra{-\alpha_*}
\end{align}
yields an upper bound on the nonclassical distance of $\ket{\psi_\pm}$.

For the even coherent state, the $Q$ function-based bounds, together with the upper bounds from Eqs.~\eqref{eq:scubb}-\eqref{eq:scuba} are shown in Fig.~1. The $Q$-function upper bound \eqref{eq:genstateub} is tighter than the upper bound corresponding to $\sigma_\beta$ for $\beta \lesssim 0.7$, while the latter coincides with the lower bound \eqref{eq:purestlb} for all practical purposes if $\beta \gtrsim 1.2$. Indeed, the maximizer $\alpha_*$ of the $Q$-function equals zero for $\beta \leq 1$, so that the lower bound cannot be tight in this regime, as explained in Section~\ref{sec:attainlb}. For $\beta >1$, the two maximizers at $\pm \alpha_*$ have $\alpha_* < \beta$ but approach $\pm \beta$ very rapidly (see inset to Fig.~1 -- this was also noted in ref.~\cite{MB03}), so that the conditions in Section~\ref{sec:attainlb} are very nearly satisfied. The upper bound \eqref{eq:scubb} shows that no matter how ``macroscopic'' the amplitude $\beta$ gets, the nonclassical distance of $\ket{\psi_+}$ approaches a maximum of 1/2. In comparison, the single-photon state has the greater nonclassical distance $1-e^{-1} \simeq 0.6321$. 

For the odd coherent state, the $Q$ function-based lower bound, together with the upper bounds from Eqs.~\eqref{eq:scubb}-\eqref{eq:scuba} as well as the distance to the uniformly phase-randomized state $\overset{\circ}{\sigma}_{\alpha_*^2}$ are shown in Fig.~2.  The $Q$-function upper bound (not shown) is looser than one of the bounds shown at all values of $\beta$.  In the limit of $\beta \rightarrow 0$, the lower bound coincides with the distance to $\overset{\circ}{\sigma}_{\alpha_*^2}$, as it should since the state approaches the one-photon number state with nonclassical distance $1-e^{-1}$. For $\beta \gtrsim 0.5$, $\ket{\psi_-}$  is closer to $\sigma_{\alpha_*}$ than to $\overset{\circ}{\sigma}_{\alpha_*^2}$. The upper bounds from $\sigma_{\alpha_*}$ and $\sigma_{\beta}$ reach the asymptotic value $1/2$ of the lower bound for all practical purposes for $\beta \gtrsim 1.5$. As for the even coherent state, this is because the maximizer $\alpha_* \rightarrow \beta$ for large $\beta$, leading to the conditions for saturation of the $Q$-function lower bound being very nearly satisfied.  From the bounds presented here, we see that the nonclassical distance of the odd coherent state is bounded above by $\sim 0.65$ for any value of $\beta$ and appears to be a decreasing function of $\beta$.

\begin{figure}[t] \label{fig:ecatstate}
\centering\includegraphics[trim=82mm 65mm 88mm 72mm, clip=true,width=\columnwidth]{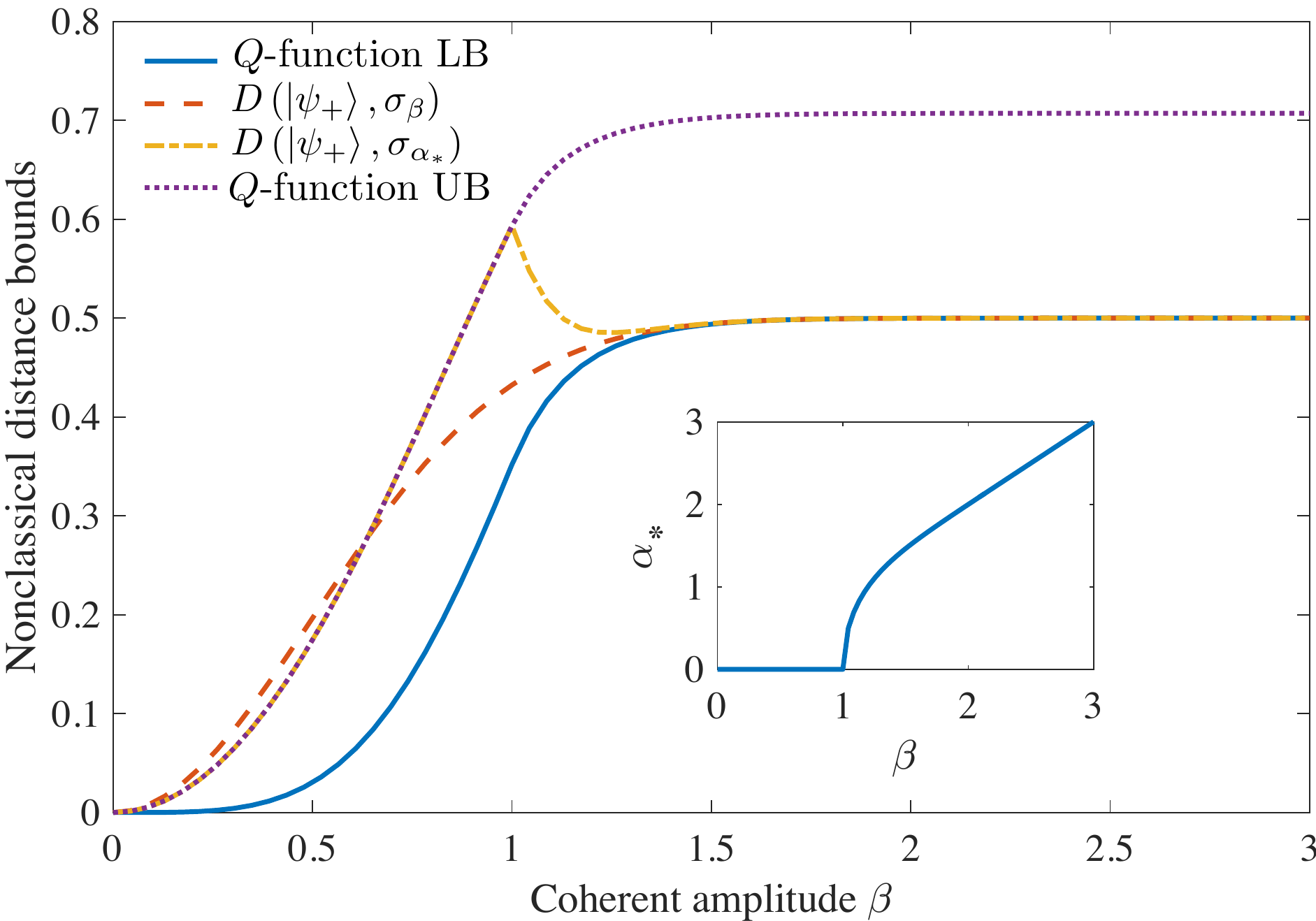}
%\onecolumngrid
\caption{(Color online) \textbf{Bounds on the nonclassical distance of the even coherent state:} The $Q$-function-based lower (solid curve) and upper (dotted curve) bounds on the nonclassical distance (Eqs.~\eqref{eq:2smb}) of the even coherent state $\ket{\psi_+}$ as a function of the coherent amplitude $\beta$. The achievable trace distances to the classical states of Eq.~\eqref{eq:scsigma} (dashed curve) and Eq.~\eqref{eq:scuba} (dash-dotted curve). \textbf{ Inset:} The maximizer $\alpha_* \geq 0$ of the $Q$ function as a function of $\beta$.} 
\end{figure}

Mixing the states \eqref{eq:scstates} with the vacuum at an $\eta: 1-\eta$ beam splitter gives rise to a two-mode \emph{entangled coherent state} \cite{San12}
\begin{align} \label{eq:ecs}
\ket{\psi'_\pm} = \frac{\cket{\sqrt{\eta}\,\beta}\cket{\sqrt{1-\eta}\,\beta} \pm\cket{-\sqrt{\eta}\,\beta}\cket{-\sqrt{1-\eta}\,\beta}}{\sqrt{2\cl{N}_\pm}}.
\end{align}
If $\sqrt{\eta}\,\beta \ll 1$ and $\sqrt{1-\eta}\,\beta \gg 1$, the state exhibits micro-macro entanglement in the spirit of the Schr\"odinger-cat thought experiment and has also been realized experimentally \cite{LGC+13}. Linear optics invariance dictates that the state \eqref{eq:ecs} has the same nonclassical distance as the original single-mode cat state regardless of the value of $\eta$. The  degree of entanglement of entangled coherent states has been studied by several authors \cite{HvEN+01,*vEH01,HDBK+16}.  Since entanglement between the output modes of a beam splitter is closely related to nonclassicality at its input \cite{ACR05,JLC13,SCE+11}, the quantitative relations between nonclassical distance of the input state and the entanglement entropy at the output merit further investigation.

\begin{figure}[t] \label{fig:ocatstate}
\centering\includegraphics[trim=78mm 65mm 90mm 72mm, clip=true,width=\columnwidth]{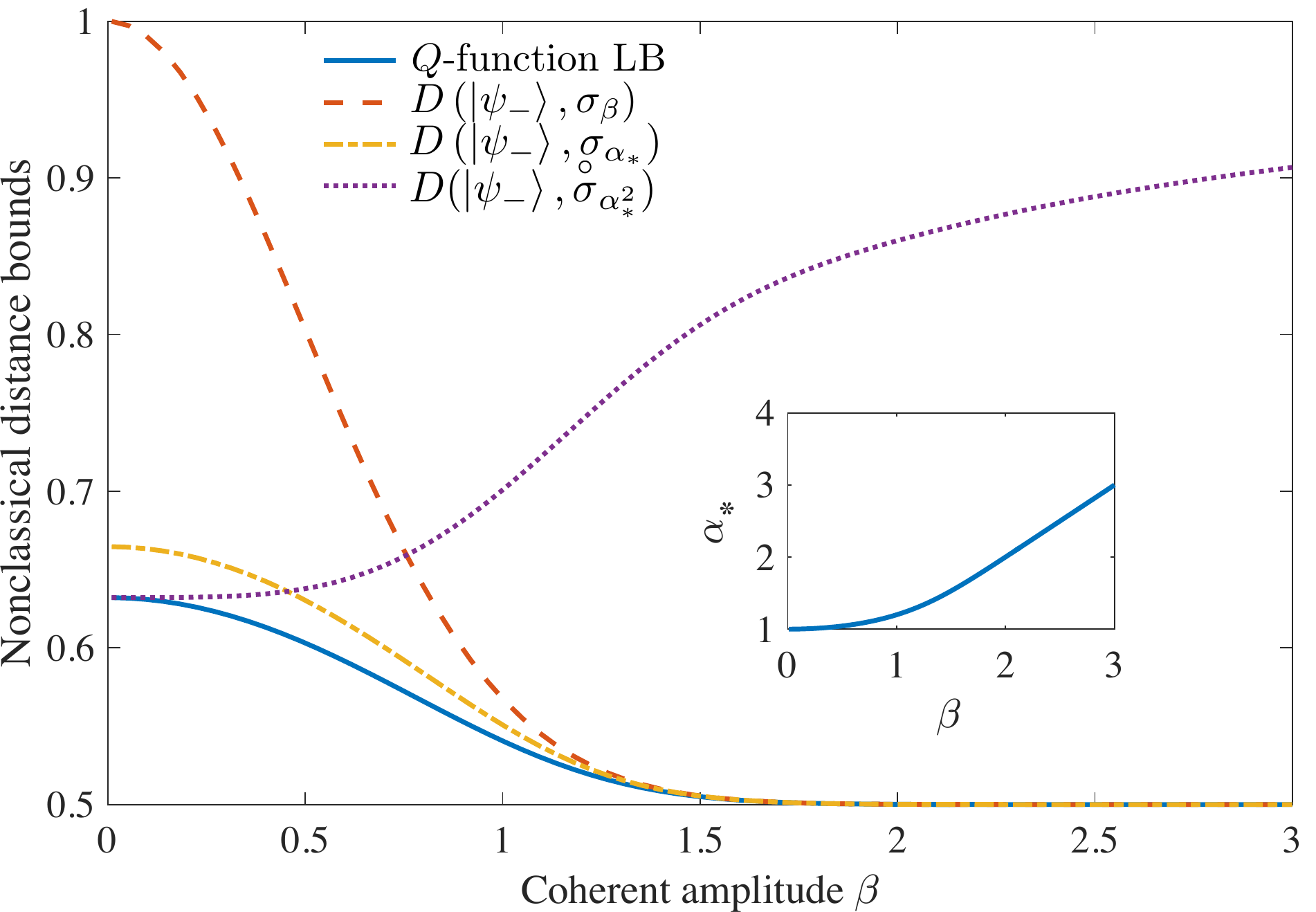}
%\onecolumngrid
\caption{(Color online) \textbf{Bounds on the nonclassical distance of the odd coherent state:} The $Q$-function-based lower bound (solid curve) on the nonclassical distance of the odd coherent state $\ket{\psi_-}$ as a function of the coherent amplitude $\beta$. The achievable trace distances to the classical states of Eq.~\eqref{eq:scsigma} (dashed curve), Eq.~\eqref{eq:scuba} (dash-dotted curve) and the phase-randomized coherent state \eqref{eq:prcs} with mean energy $\alpha_*^2$ (dotted curve). \textbf{ Inset:} The maximizer $\alpha_* > 0$ of the $Q$ function as a function of $\beta$.} 
\end{figure}
\subsection{Mixture of vacuum and number state}

As a final example, consider the state
\begin{align} \label{eq:numvacmixture}
\rho = (1-\eta)\ket{0}\bra{0}{} + \eta \ket{n}\bra{n}
\end{align}
that mixes a vacuum component with a number-state component with $n \geq 1$, and is nonclassical for any $\eta>0$ because it has zero probability of counting $m$ photons for $m \neq 0,n$. The nonclassicality of the $n=1$ case was extensively studied in \cite{LS02}, while the nonclassical depth of this state was considered in \cite{MB03}. Convexity of the nonclassical distance gives the upper bound
\begin{align} \label{eq:conub}
\delta(\rho) \leq \eta \,\delta(\ket{n}) = \eta \,(1-\gamma_n).
\end{align}
The lower bound \eqref{eq:mixedstatelb} is hard to compute, but since $D(\rho, \ket{n}) = 1 -\eta$, we can use \eqref{eq:tribound} to get
\begin{align} \label{eq:trilb}
\eta - \gamma_n \leq \delta(\rho),
\end{align}
which is useful if $\eta > \gamma_n$. These upper and lower bounds are shown in Fig.~3.

Using an argument similar to that in Sec.~\ref{sec:clpchannels}, we can say something about the form of a classical state $\sigma$ that satisfies $\delta(\rho) = D(\rho,\sigma)$. Consider the quantum channel $M:S \rightarrow S$ corresponding to making an ideal measurement in the number basis  that maps a state $\tau$ into
\begin{align} \label{eq:measmap}
M\tau &:= \sum_{n=0}^\infty \ket{n}\bra{n} \tau \ket{n} \bra{n} \\
& = \frac{1}{2\pi} \int_0^{2\pi} \diff\theta \,e^{-i\theta a^\dag a} \,\tau \, e^{i\theta a^\dag a}.
\end{align}
The last equation shows that $M$ can be implemented as a randomized phase shift over $[0,2\pi]$ and is hence  classicality-preserving. We also have $M\rho =\rho$. Therefore, for any $\sigma \in S_{\rm cl}, 
D(\rho,\sigma) \geq D(M\rho,M\sigma) = D(\rho, M\sigma)$ with $M\sigma \in S_{\rm cl}$. Therefore, it suffices to restrict the optimization to all classical states diagonal in the number basis, i.e., those with a circularly symmetric $P$ function. However, this latter optimization appears to be non-trivial and may require a numerical approach.

\begin{figure}[t] \label{fig:vacnmix}
\centering\includegraphics[trim=84mm 64mm 86mm 73mm, clip=true,width=\columnwidth]{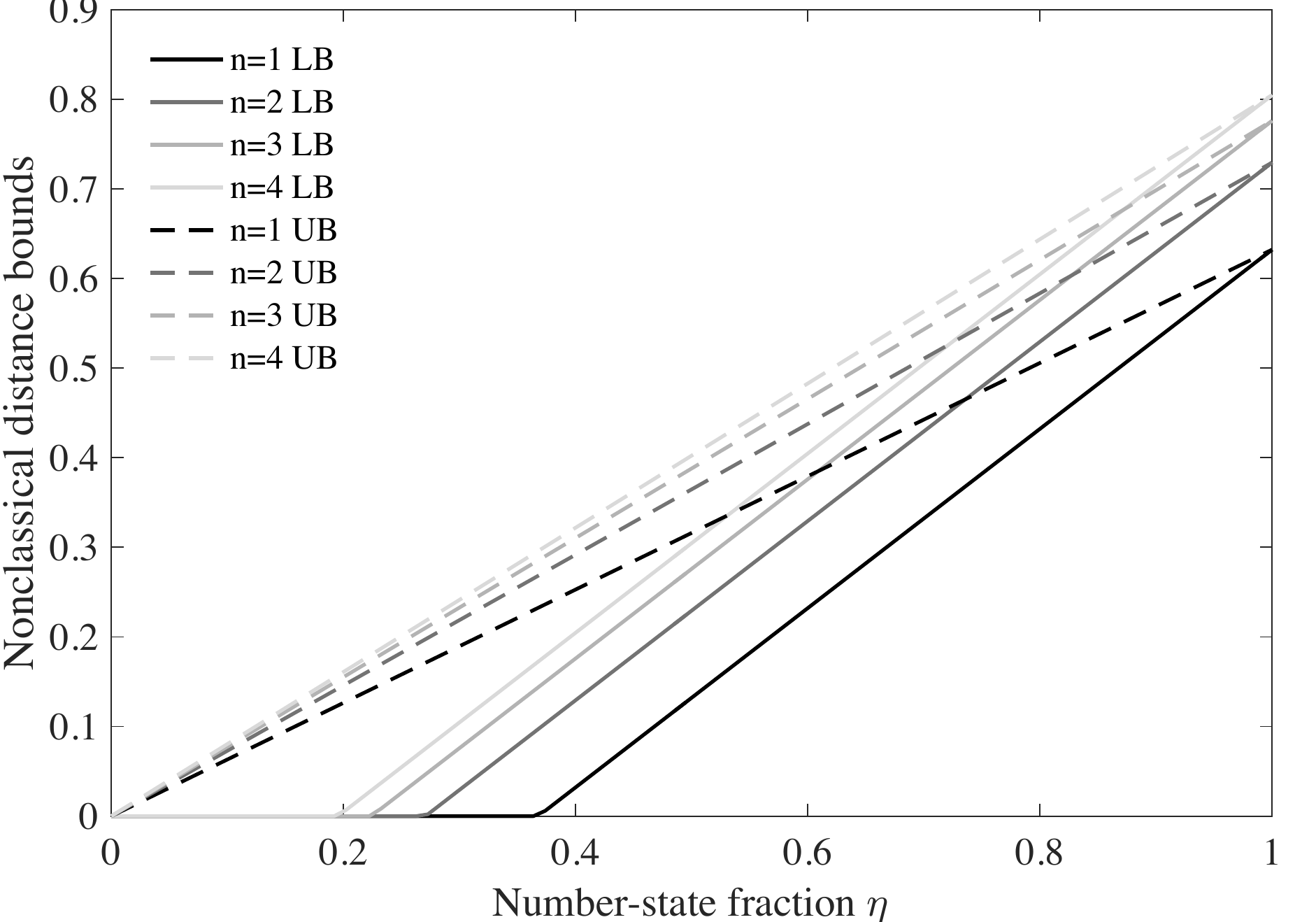}
%\onecolumngrid
\caption{The lower (solid curves) and upper (dashed curves) bounds (Eqs.~\eqref{eq:trilb} and \eqref{eq:conub}) on the nonclassical distance of the number-state-vacuum mixture \eqref{eq:numvacmixture} as a function of  $\eta$ for $n=1, 2, 3, 4$. Grayscale shading becomes lighter as $n$ increases.} 
\end{figure}
\section{Discussion and Outlook} \label{sec:discussion}
We have revisited the nonclassical distance defined in \cite{Hil87} in a multimode setting, studied its properties, developed new bounds on it, and elucidated the cases for which our Husimi-function lower bound is tight for pure states. The number states, multimode single-photon states, and multimode N00N states constitute, to the best of our knowledge, the first examples for which the nonclassical distance has been calculated exactly. Further work is needed to verify if our lower or upper bounds are tight for other important nonclassical states. Our Husimi-function lower bound \eqref{eq:purestlb} can be used to show that the nonclassical distance of the one-mode (two-mode) squeezed vacuum states can be made arbitrarily close to unity by increasing the degree of squeezing (entanglement). However, as mentioned in Sec.~\ref{sec:attainlb}, the lower bound cannot be tight for pure multimode Gaussian states. In view of the practical importance of these states, it would be useful to get good estimates of their nonclassical distance.

In ref.~\cite{Hil89}, upper bounds on the nonclassical distance of a state in terms of its total noise \cite{Sch86} or average energy were derived. In view of the interplay between the photon number and the number of modes in determining the nonclassical distance for some of our examples, it would be interesting to seek upper bounds on the nonclassical distance of a state in terms of its total average energy and the number of modes $M$. 

The generation of large-amplitude optical Schr\"odinger-cat states is of great interest from the viewpoints of both fundamental physics and applications such as optical quantum computation, with states of increasing amplitudes being generated in recent years \cite{OJT+07,LGC+13,HLJ+15}. Since the generation of large-amplitude coherent-state superpositions appears to be challenging, it is somewhat surprising that the nonclassical distance of the  cat states is bounded away from unity regardless of the superposition amplitude. Further study is required to see if this indicates that alternative preparation strategies for large-amplitude cat states exist.

The attenuator and additive Gaussian noise channels preserve classicality and degrade nonclassicality \cite{ACR05,SVL06,YN09,SIS11}. In view of their ubiquity, it would be very useful to study quantitatively how the nonclassical distance degrades at the output of such channels.

\iffalse
\begin{align}
 \left\{
	\begin{array}{ll}
		\frac{1}{\sqrt{T}} \exp\lp-i \frac{2\pi mt}{T}\rp  & \mbox{if } t \in[-T/2,T/2]  \\
		0 & \mbox{otherwise,}
	\end{array}
\right.
\end{align}
\fi

\section{Acknowledgements}
I thank Mark Hillery, Krishna Kumar Sabapathy, and an anonymous referee for useful comments. This work is supported by the Singapore National Research Foundation under NRF Grant No.~NRF-NRFF2011-07 and the Singapore Ministry of Education Academic Research Fund Tier 1 Project R-263-000-C06-112.


%\section*{Supplementary Material}
\end{document}